\documentclass[11pt]{article}
\usepackage{graphicx} 
\usepackage{cite}

\mathcode`O="724F

\newtheorem{theorem}{Theorem}

\newtheorem{observation}[theorem]{Observation}     

\begin{document}
\title{A Steady State Model for Graph Power Laws}
\author{David Eppstein\thanks{Dept. Inf. \& Comp. Sci.,
UC Irvine, CA 92697-3425, USA,
{\tt\{eppstein,josephw\}@ics.uci.edu}.}
\and Joseph Wang$^*$}        

\date{ }
\maketitle    

\begin{abstract}
Power law distribution seems to be an important characteristic
of web graphs. Several existing web graph models generate power law graphs by
adding new vertices and non-uniform edge connectivities to 
existing graphs.
Researchers have
conjectured that preferential connectivity and incremental growth 
are both required for the power law distribution.
In this paper, we propose a different web graph model with
power law distribution that does not require incremental growth.
We also provide a comparison of our model with several others in 
their ability to predict web graph clustering behavior.
\end{abstract}

\section{Introduction}
The growth of the World Wide Web (WWW) has been explosive and
phenomenal. 
Google~\cite{google} has more than
$2$ billion pages searched as of February $2002$. 
The Internet Archive~\cite{archive} has
$10$ billion pages archived as of March $2001$. 
The existing growth-based models~\cite{ACL2000,BA1999,KRRSTU2000} are adequate to explain
the web's current graph structure. 
It would be interesting to know
if a different model will be needed as the web's growth rate slows
down~\cite{oclc} 
while its link structure continues to evolve.

\subsection{Why Power Laws?}
Barab\'asi et al.~\cite{BAJ1999,BAJ2000}
and Medina et al.~\cite{MMB2000}
stated that preferential connectivity and incremental growth 
are both required for the power law distribution observed in
the web. The importance of the preferential connectivity
has been shown by several researchers~\cite{BA1999,HA1999}.

Faloutsos et al.~\cite{FFF1999} observed that the internet
topology exhibits power law distribution in the form of
$y = x^\alpha$. 
When studying web characteristics, the documents can be viewed as
vertices in a graph and the hyper-links as edges between them.
Various researchers~\cite{AJB1999,BA1999,KKRRT1999,KRRT1999}
have independently showed the power law distribution in the 
degree sequence of the web graphs. 
Huberman and Adamic~\cite{AH2000,HA1999} showed a power law distribution
in the web site sizes.  See~\cite{KL2001} for a summary
of works on web graph structure.

Medina et al.~\cite{MMB2000} showed that topologies generated
by two widely used generators, the Waxman model~\cite{Waxman88},
and the GT-ITM tool~\cite{CDZ1997}, do not have power law distribution 
in their degree sequences. Palmer and Steffan~\cite{PS2000} proposed a 
power law degree generator that recursively partitions
the adjacency matrix into an $80$-$20$ distribution. However,
it is unclear if their generator actually emulates
other web properties.

The power law distribution seems to be an ubiquitous 
property. The power law distribution occurs in 
epidemiology~\cite{RA96},
population studies~\cite{PalmerWhitge94}, 
genome distribution~\cite{HN97,QLG2001}, 
various social phenomena~\cite{BAJB2000,OS2001},
and massive graphs~\cite{ABW98,ACL2000}. For the power
law graphs in biological systems, the connectivity
changes appear to be much more important than growth in 
size due to the long time-scale of biological evolution.

\subsection{Properties for Graph Model Comparison}
Another important web graph property that has been looked at is diameter. However, there are conflicting results
in the published papers.
Albert et al.~\cite{AJB1999} stated that web graphs
have the {\sl small world phenomenon}~\cite{Milgram67,Watts1999},
in which the diameter $\Delta$ is roughly $0.35 + 2.06 \lg n$,
where $n$ is the size of the web graph. For $n = 8 \times 10^8$,
$\Delta \approx 19$. 
Lu~\cite{Lu2001} proved
the diameters of random power law graphs are logarithmic 
function of $n$ under the model proposed by Aiello et al.~\cite{ACL2000}.
However, Broder et al.~\cite{BKMRRSTW2000} 
showed that, over $75\%$ of the time, there is no directed path between
two random vertices. If there is a path, the average distance
is roughly $16$ when viewing web graph as directed graph
or $6.83$ in the undirected case. 

Currently, there are few theoretical graph models
\cite{ACL2000,BA1999,KRRSTU2000,PS2000} for generating power law graphs.
There are very few comparative studies that would allow us to determine
which of these theoretical models are more accurate models of the web. 
We only know that the model proposed by Kumar et al.~\cite{KRRSTU2000} generates more bipartite cliques
than other models. They believe clustering to be an
important part of web graph structures that was
insufficiently represented in previous models~\cite{ACL2000,BA1999}.

\subsection{New Contributions}
In this paper, we show power law graphs do not require 
incremental growth, by developing a graph model which
(empirically) results in power laws by evolving
a graph according to a Markov process while maintaining
constant size and density.

We also describe an easily computable graph property that can be 
used to capture cluster information in a graph without
enumerating all possible subgraphs.  We use this property to compare our model with others and with actual web data.

\section{Steady State Model}
Our $Steady State$ (SS) model is very simple in comparison with 
other web graph
models~\cite{ACL2000,BA1999,KRRSTU2000,PS2000}. It consists of
repeatedly removing and adding edges in a sparse random
graph $G$. 

Let $m$ be $\Theta(n)$.  We generate an initial sparse random graph
$G$ with $m$ edges and $n$ vertices, by randomly adding edges between
vertices until we have $m$ edges.  As discussed below, the initial random distribution of edges is unimportant for our model.

We then iterate the following steps $r$ times on $G$,
where $r$ is a parameter to our model.
\begin{enumerate}
\item Pick a vertex $v$ at random. If there is no edge
incident upon $v$, we repeat this step until $v$ has nonzero degree.
\item Pick an edge $(u, v) \in G$ at random. 
\item Pick a vertex $x$ at random.
\item Pick a vertex $y$ with probability proportional to degree.
\item If $(x, y)$ is not an edge in $G$ and $x$ is not equal to $y$, 
then remove edge
$(u, v)$ and add edge $(x, y)$.  
\end{enumerate}
One can view our model as an aperiodic Markov chain with some limiting 
distribution. If we repeat the above steps long enough, the
random graphs generated by this model will be close to this limiting distribution, no matter what
the initial random sparse graph is. 
Note that unlike other models~\cite{ACL2000,KRRSTU2000},
the graphs generated by our model do not contain self-loops nor 
multiple edges between two vertices.

Barab\'asi et al.~\cite{BAJ1999} also proposed a non-growth model,
which failed to produce a power law distribution. 
Both models have preferential connectivity features.
However, there are several
differences between our model and theirs. First, our edge
set is fixed and the initial graph is generated via classical 
random graph models~\cite{ER59,JLR2000}. Second, our
model has ``rewiring'' feature similar to one in
the small world model~\cite{BAJ1999,Milgram67,Watts1999}.

\subsection{Simulation Results}
\begin{figure}
\centerline{\includegraphics[height=3in]{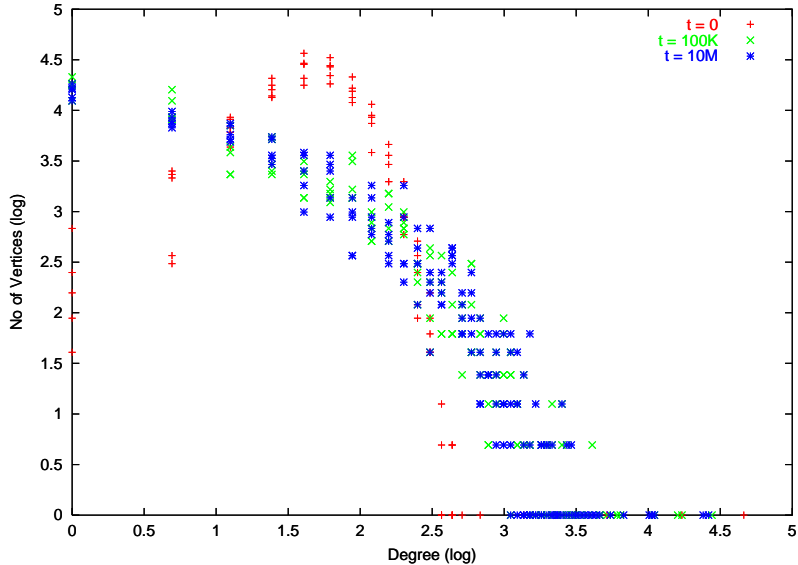}}
\caption{Initial $G(500, 1500)$, \& $G$ After $100K$ and $10M$ Steps}
\label{fig:v500_1500}
\end{figure}

\begin{figure}
\centerline{\includegraphics[height=3in]{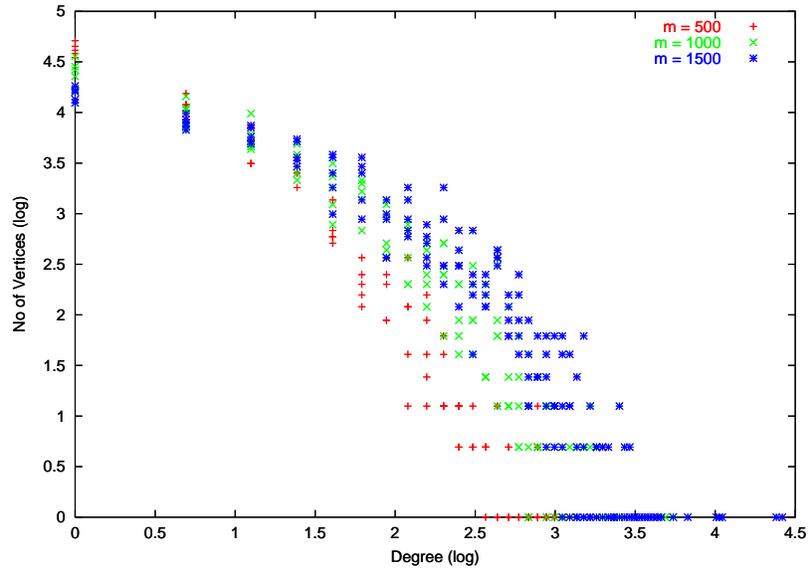}}
\caption{$G(500, 500)$, $G(500, 1000)$, and $G(500, 1500)$ After $10M$ Steps} 
\label{fig:v500_d}
\end{figure}

\begin{figure}
\centerline{\includegraphics[height=3in]{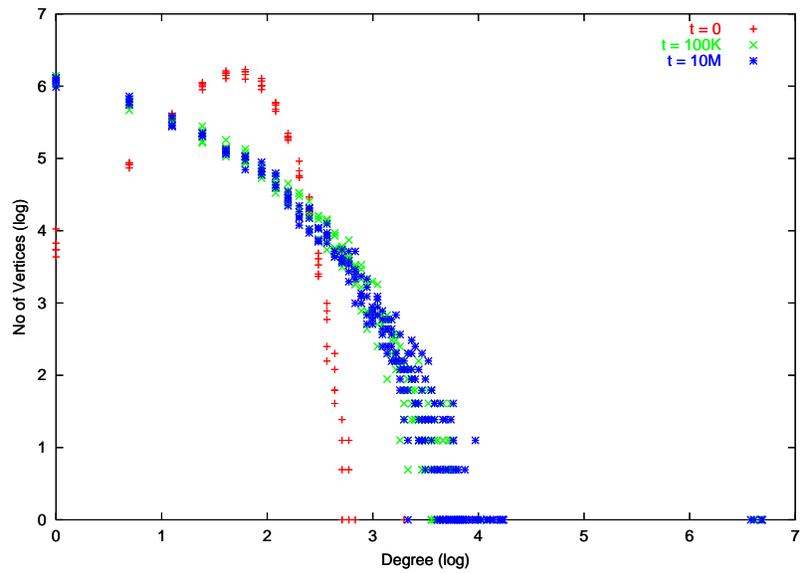}}
\caption{Initial $G(3000, 9000)$, \& $G$ After $100K$ and $10M$ Steps}
\label{fig:v3000_9000}
\end{figure}

\begin{figure}
\centerline{\includegraphics[height=3in]{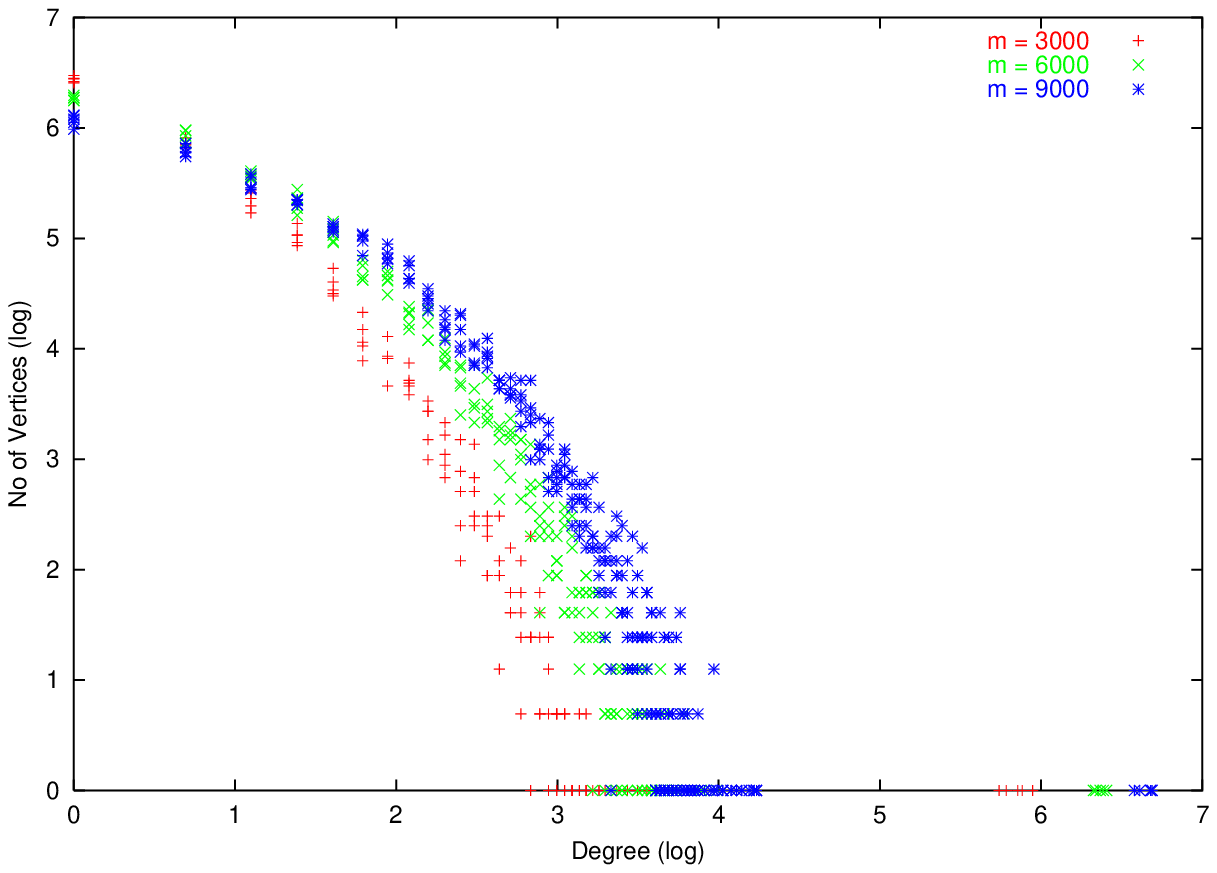}}
\caption{$G(3000, 3000)$, $G(3000, 6000)$, and $G(3000, 9000)$ After $10M$ Steps} 
\label{fig:v3000_d}
\end{figure}

We simulated our model on graphs of different 
sizes, $(500 \le n \le 5000)$, and 
densities $m \over n$, $(1 \le {m \over n} \le 3)$.
We repeated each simulation $5$ times, and performed $r = 10000000$ edge
deletion/insertion operations on each graph.
The vertices' degree distributions appear to converge to
power law distributions as the number of edge 
deletion/insertion operations increases. 
Some of our simulation results are shown in 
Figures ~\ref{fig:v500_1500} - ~\ref{fig:v3000_d}.
Figures ~\ref{fig:v500_1500} and ~\ref{fig:v3000_9000}
show degree distributions at
various stage of simulations. 
Figures ~\ref{fig:v500_d} and ~\ref{fig:v3000_d}
show degree distributions for graphs with different
densities $m \over n$.

\smallskip

\smallskip

\smallskip

\smallskip

\section{Cluster Information}
Given a subgraph $S$ of $G$, $d_S(v)$ is the 
degree for vertex $v$ in $S$.
Here we examine the maximum degree $d_{\mathrm max}$ in all subgraphs, 
which is defined as
$$d_{\mathrm max} = {\mathrm max}_S \ \ {\mathrm min}_{v \in S} \ \ d_S(v).$$
We use $d_{\mathrm max}^{M}$ to denote the value obtained under graph
model $M$.

\begin{figure}
\centerline{\includegraphics{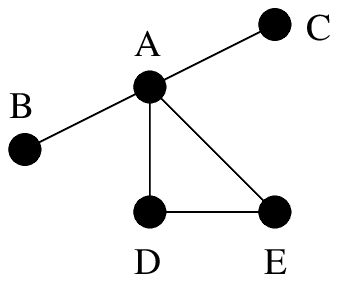}}
\caption{Minimal Degree Vertex Elimination} \label{fig:mindegree}
\end{figure}

To compute $d_{\mathrm max}$ for a graph $G$, we perform
the following steps until $G$ becomes empty:
\begin{enumerate}
\item Select a minimum degree vertex $v$ from $G$.
\item Set $d_{\mathrm max}$ to $d(v)$ if $d(v) > d_{\mathrm max}$.
\item Remove vertex $v$ and its edges from $G$. 
\end{enumerate}
The above steps correctly compute $d_{\mathrm max}$ because
we cannot remove any vertices of $S$ until the degree
of the current subgraph reaches $d_{\mathrm max}$.
The minimal degree elimination sequence
for graph in Figure ~\ref{fig:mindegree} will be $B, C, A, D,$ and $E$. 
The degrees when those vertices got eliminated are
$1, 1, 2, 1,$ and $1$. $d_{\mathrm max}$ is $2$ since 
${\mathrm max} \{ 1, 1, 2, 1, 1\} = 2$ .

\begin{observation}
For any model $M$ that constructs a graph by adding a vertex at a time,
and for which each newly added vertex has the same degree
$d = {m \over n}$, $d_{\mathrm max}^{M} = d$.
\end{observation}

Thus the Barab\'asi and Albert model (BA)~\cite{BA1999} or the linear growth
copying model in~\cite{KRRSTU2000} have the same value for
$d_{\mathrm max}$ for graphs of all sizes once $d ={m \over n}$ is fixed.

\begin{observation}
The web graph generated by the linear model has 
minimum vertex degree of $d = {m \over n}$.
\end{observation}

Hence, the linear model may not encapsulate all the crucial properties
in a web graph if there are significant numbers of vertices with
degree less than ${m \over n}$.

\subsection{Web Crawl and Simulation Data}
We performed a web crawl on various Computer Science
department web sites.
We then used the $ACL$ model~\cite{ACL2000} to generate
new graphs from degree sequences in the actual web
graphs. We also ran the $SS$ model using $n$ and $m$ values
from the actual web graphs with $10000000$
edge insertion/deletion steps. 
For each graph, we run both models $5$ times.
The following table
shows the means $\mu$ and the
standard deviations $\sigma$ for 
$d_{\mathrm max}$ values
using the $ACL$ model and the $SS$ model.

\begin{center}
\begin{tabular}{|l|c|c|c|c|c|c|c|} \hline
Site & $n$ & $m$ & $d_{\mathrm max}$ & 
$\mu_{ACL}$ &  
$\sigma_{ACL}$ &
$\mu_{SS}$ &  
$\sigma_{SS}$ \\ \hline
arizona & 5315 & 16892 & 15 & 10 & 0 & 8 & 0 \\ \hline
berkeley & 2826 & 22957 & 45 & 21.6 & 0.547 & 16 & 0 \\ \hline
caltech & 622 & 4830 & 7 & 5.8 & 0.447 & 12.8 & 0.447 \\ \hline
cmu & 2052 & 23821 & 57 & 37.2 & 0.447 & 20 & 0.707 \\ \hline
cornell & 7145 & 14919 & 17 & 19.4 & 0.547 & 6 & 0 \\ \hline
harvard & 915 & 9327 & 21 & 12.6 & 0.894 & 16.4 & 0.547 \\ \hline
mit & 4861 & 15360 & 31 & 24.4 & 0.547 & 7 & 0 \\ \hline
nd & 1913 & 16328 & 33 & 29.2 & 0.447 & 15.4 & 0.547 \\ \hline
stanford & 2553 & 25693 & 27 & 14.6 & 0.547 & 18.4 & 0.547 \\ \hline
ucla & 2718 & 19755 & 22 & 16.6 & 0.547 & 14.2 & 0.447 \\ \hline
ucsb & 5236 & 10338 & 22 & 13.8 & 0.447 & 5 & 0 \\ \hline
ucsd & 553 & 3885 & 15 & 7.2 &0.447 & 11.8 & 0.447 \\ \hline
uiowa & 1410 & 12258 & 8 & 8.8 & 0.447 & 15.2 & 0.447 \\ \hline
uiuc & 5623 & 28872 & 29 & 21 & 0 & 11.8 & 0.836 \\ \hline
unc & 1465 & 5446 & 17 & 9.8  & 0.447 & 8 & 0 \\ \hline
washington & 7001 & 24901 & 17 & 12 & 0 & 9 & 0 \\ \hline
\end{tabular}

Table $1$: $d_{\mathrm max}$ from Actual Web Crawl and Model Simulation 
\end{center}
\smallskip

In general, the $ACL$ model and the $SS$ model are
generating less clustered graphs
than what we see on actual web graphs.
This implies that we need a more detailed model of web
graph clustering behavior.

\section{Conclusion and Open Problems}
Previously, researchers have conjectured that preferential
connectivity and incremental growth are necessary factors
in creating power law graphs. 
In this paper, we provide a model of graph evolution that
produces power law without growth. Our $Steady State$ model 
is very simple in comparison with other graph models
\cite{KRRSTU2000}. It also does not require prior degree
sequences as in the $ACL$ model~\cite{ACL2000}.

The difficulty in comparing various models~\cite{ACL2000,BA1999,KRRSTU2000} is that
each model has different parameters and inputs.
Here we provide a simple graph property
$d_{max}$
 that captures 
the clustering behavior of graphs without complicated 
subgraph enumeration algorithm.
It can be useful in gauging the accuracy of various models.

From our web crawl data, we know that the linear models 
such as Barab\'asi's~\cite{BA1999} 
are not the best ones to use when considering
$d_{\mathrm max}$.
Both $ACL$ and $SS$ models are
not generating dense-enough subgraphs when
comparing against the actual web graphs.
Thus, we need a better web graph model
that mimics actual web graph clustering behavior.

Here are some of our open problems:
\begin{enumerate}
\item Can one prove theoretically
that the SS method actually has a power law distribution?
\item How long does it take for our model
to reach a steady state?
\item What are other simple web graph properties
that we can use to determine the accuracy of
various models?
\item Are there any technique such as graph products
that we can use to generate realistic massive web graphs in
relatively short times?
\end{enumerate}

\bibliography{bibdata}
\bibliographystyle{acm}  
\vfill
\end{document}